\documentclass[10pt]{iopart}

\usepackage{amssymb}
\usepackage{pstricks}
\usepackage{epsfig}
\usepackage{color}

\def\<{\left<}
\def\>{\right>}
\def\ket|#1>{\left|#1\right>}
\def\bra<#1|{\left<#1\right|}
\def\elem<#1|#2|#3>{\left<#1\right|#2\left|#3\right>}
\def\({\left(}
\def\){\right)}

\def\R{{\mathbb R}}
\def\pl{\partial}
\psset{unit=1mm}

\begin{document}

\title[Random geometry and the KPZ universality class]
{Random geometry and the Kardar-Parisi-Zhang universality class}

\author{Silvia N. Santalla$^1$, Javier Rodr\'{\i}guez-Laguna$^{2,3}$,
  Tom LaGatta$^4$ and Rodolfo Cuerno$^2$}
\address{$^1$ Physics Department and Grupo Interdisciplinar de
  Sistemas Complejos (GISC), Universidad Carlos III de
  Madrid, Avenida de la Universidad 30, 28911 Legan\'es, Spain}
\address{$^2$ Mathematics Department and GISC, Universidad Carlos III
  de Madrid, Avenida de la Universidad 30, 28911 Legan\'es, Spain}
\address{$^3$ ICFO-Institute of Photonic Sciences, Barcelona, Spain}
\address{$^4$ Courant Institute, New York, United States}

\date{February 6, 2015}

\begin{abstract}
We consider a model of a quenched disordered geometry in which a
random metric is defined on $\R^2$, which is flat on average and
presents short-range correlations. We focus on the statistical
properties of balls and geodesics, i.e., circles and straight
lines. We show numerically that the roughness of a ball of radius $R$
scales as $R^\chi$, with a fluctuation exponent $\chi \simeq 1/3$,
while the lateral spread of the minimizing geodesic between two points
at a distance $L$ grows as $L^\xi$, with wandering exponent value
$\xi\simeq 2/3$. Results on related first-passage percolation (FPP)
problems lead us to postulate that the statistics of balls in these
random metrics belong to the Kardar-Parisi-Zhang (KPZ) universality
class of surface kinetic roughening, with $\xi$ and $\chi$ relating to
critical exponents characterizing a corresponding interface growth
process. Moreover, we check that the one-point and two-point
correlators converge to the behavior expected for the Airy-2 process
characterized by the Tracy-Widom (TW) probability distribution
function of the largest eigenvalue of large random matrices in the
Gaussian unitary ensemble (GUE). Nevertheless extreme-value statistics
of ball coordinates are given by the TW distribution associated with
random matrices in the Gaussian orthogonal ensemble. Furthermore, we
also find TW-GUE statistics with good accuracy in arrival times.
\end{abstract}




\section{\label{introduction}Introduction}

Random geometry is a branch of mathematics \cite{Adler} with deep
connections to physics, ranging from statistical mechanics to quantum
gravity \cite{Itzykson_Drouffe,Booss.book}. For example, thermal
fluctuations of important biophysical objects, like fluid membranes,
can be naturally accounted for through the framework of random
geometry \cite{Nelson_et_al,Boal}. The effect of thermal or quantum
fluctuations of the geometry on systems featuring strong correlations,
such as those underlying a continuous phase transition, is typically
relevant, in the sense that they modify the values of the critical
exponents \cite{Ambjorn_97}. For 2D systems, this modification is
governed by the celebrated Knizhnik-Polyakov-Zamolodchikov equations
\cite{Knizhnik_88}. If, instead of thermal or quantum fluctuations, we
consider quenched disorder in the geometry, one is naturally led to
the study of models like {\em first passage percolation} (FPP)
\cite{Hammersley_65,Howard_04}. In this discrete model, each link of a
regular lattice is endowed with a random {\em passage time}. FPP
theory studies the probability distribution of traveling times between
pairs of lattice points. Alternatively, minimal traveling times can be
regarded as distances, thereby defining a random metric. Being a
generalization of the Eden model \cite{Howard_04,Kesten_03}, FPP has
played an important role in statistical physics, as an important step
for the analysis of other interacting particle systems like the
contact process or the voter model. More recently, additional
interest in the model derives from its properties when defined on
realistic (disordered) graphs \cite{Eckhoff_13}, such as those
occurring in e.g.\ communications or economic systems \cite{Newman}.

Inspired by studies in FPP, recent works have dealt with geodesics and
balls in a two-dimensional plane endowed with suitable random metrics
\cite{Lagatta_10,Lagatta_14}. By suitable, we mean that the metric is
on average flat and presents only short-range correlations. In other
terms, the geometric properties are considered over distances much
larger than either the curvature radius or the correlation
lengths. The {\em geodesics} on these random manifolds present many
interesting properties. Let us consider two points which are separated
by an Euclidean distance $L$. The minimizing geodesic on the random
metric which joins them can be regarded as a random curve, when viewed
from the Euclidean point of view. Its maximal deviation from the
Euclidean straight line grows as $L^\xi$. It is also possible to study
{\em balls} on these random metrics. The ball of radius $R$ around any
point will be also a random curve, from the Euclidean point of
view. For large $R$, the shape of this curve can be shown to approach
a circumference, whose radius is proportional to $R$. It is
conjectured to lie within an annulus whose width grows as $R^\chi$
\cite{Lagatta_10,Lagatta_14}.

The so-called wandering and fluctuation exponents, $\xi$ and $\chi$,
for the geodesic and ball fluctuations, respectively, denote a certain
universal fractal nature of straight lines and circles on a random
geometry. Actually, they also occur for FPP on a lattice, where they
are known to correspond, through an appropriate interpretation
\cite{Krug_92}, to those characterizing the dynamics of a growing
interface. Basically, the boundary of a FPP ball can be thought of as
an interface which, in the wider context of models of surface kinetic
roughening \cite{Barabasi,Krug_97}, is expected to grow irreversibly,
in competition with time-dependent fluctuations and smoothing
mechanisms. Starting with a flat or a circular form, the interface
roughness (root-mean-square deviation around the mean interface
position) grows in time as $W(t)\sim t^\beta$. Also, the interface
fluctuations present a lateral correlation length which grows with
time as $\ell(t) \sim t^{1/z}$. The FPP values for the growth and
dynamic exponents, $\beta=1/3$ and $1/z=2/3$, respectively
\cite{Krug_92}, correspond to those of the so-called
Kardar-Parisi-Zhang (KPZ) universality class for one-dimensional
interfaces \cite{Barabasi,Krug_97,Kardar_PRL86}. Actually, a landmark
scaling relation that holds among exponents for systems within this
class, namely the so-called Galilean relation $\beta+1=2/z$, implies
through the interface interpretation $\xi\to 1/z$ and $\chi\to\beta$
\cite{Krug_92} that $\chi+1=2\xi$, which has been proved only very
recently for FPP under strong hypothesis
\cite{Chatterjee_13,Auffinger_14}. Rigorously speaking, the individual
values $\xi=2/3$ and $\chi=1/3$ remain conjectural for FPP.

In recent years, evidence has gathered, showing that systems in the
KPZ universality class do not only share the values of the scaling
exponents $\beta$ and $1/z$, but also the full probability
distribution of the interface fluctuations \cite{Takeuchi_11}, being
accurately described by an universal, stationary, stochastic process
that goes by the name of Airy process
\cite{Praehofer_02,Corwin_13}. This applies to discrete models
\cite{Corwin_12,Alves_11,Oliveira_12}, experimental systems
\cite{Takeuchi_PRL10,Takeuchi_11,Yunker_13,Nicoli_13}, and to the KPZ
equation itself \cite{Sasamoto_PRL10,Amir_CPAM11,Calabrese_11}. For
one-dimensional interfaces and within the context of simple-exclusion
processes ---and as a confirmation of a conjecture formulated in the
context of the polynuclear growth model
\cite{Praehofer_PRL00,Praehofer_PA00}--- it has been rigorously proved
that, for a band geometry, interface fluctuations follow the
Tracy-Widom (TW) probability distribution function associated with
large random matrices in the Gaussian orthogonal ensemble (GOE), while
for a circular setting they follow the TW distribution associated with
the Gaussian unitary ensemble (GUE)
\cite{Johansson_CMP00,Dotsenko_JSM10,Takeuchi_JSTAT12}. Universal
fluctuations of TW type are also known to show up in FPP systems, but
in this case the variable whose fluctuations are typically considered
is the time of arrival, rather than the radius
\cite{Johansson_00}. The values of the fluctuation and wandering
exponents in the random metric problem suggest a direct relation to
non-equilibrium processes in the KPZ universality class
\cite{Lagatta_10,Lagatta_14}.

In this work we develop an adaptive numerical algorithm to explore the
shapes of balls in arbitrary two-dimensional Riemannian manifolds, and
specialize it to work on random metrics of the desired properties. Our
algorithm is based on the one used to solve the covariant KPZ equation
\cite{LSC_JSTAT11,SLC_PRE14}. We show numerically that those balls, as
conjectured \cite{Lagatta_10,Lagatta_14}, follow KPZ
scaling. Minimizing geodesics are studied, and their fluctuations are
shown to scale in the expected way. Moreover, radial fluctuations are
shown to follow Airy-2 process statistics both in the one-point and the
two-point functions \cite{Praehofer_02,Corwin_13}. The extreme-value
statistics of ball coordinates turn out to be given nevertheless 
by the TW-GOE distribution, akin to previous experimental and theoretical 
results in circular geometries \cite{Takeuchi.JSP.12,Johansson.CMP.03}. 
Finally, we also study a related variable, the time of arrival, and show that it again follows TW-GUE statistics.

The paper is structured as follows. Section \ref{model} discusses the
basics of geometry in random metrics. The numerical algorithm is
described in section \ref{numerics}, followed by a detailed study of
balls and geodesics in section \ref{balls}. The Airy-2 statistics of
radial fluctuations is discussed in detail in section
\ref{airy}. Section \ref{timeofarrival} studies the time of arrival,
while Section \ref{conclusions} ends by presenting our conclusions and
plans for future work.


\section{Geometry in random metrics}
\label{model}

Let us consider the Euclidean plane $\R^2$, endowed with the usual
Euclidean distance, $d_E$. Let us now define a manifold ${\cal M}$
obtained when a(n almost sure) $C^{\infty}$ metric tensor field $g$ is
imposed upon $\R^2$, inducing a distance function $d_g$. Let us
consider, following \cite{Lagatta_10,Lagatta_14}, an ensemble of such
smooth metric fields which fulfill the following conditions:

\begin{itemize}
\item Independence at a distance: the metric tensor at any two points
  whose (Euclidean) distance is larger than a cutoff $r_0$ are
  independent (this implies a compactly-supported correlated
  function).
\item Statistical homogeneity and isotropy: the probability
  distribution function for the metric tensor values is invariant
  under arbitrary translations and rotations in the plane.
\item Almost sure smoothness: with probability one, the metric tensor
  is everywhere $C^\infty$-smooth.
\end{itemize}

Examples of such random metric tensors can be found in
\cite{Lagatta_14}.

The metric tensor field $g$ can be visualized as a mapping that
attaches to each point two orthogonal directions, $\vec v_1$ and $\vec
v_2$, and two metric eigenvalues, $\lambda_1$ and
$\lambda_2$. Alternatively, we can think that each point in $\R^2$
gets an ellipse attached, with principal directions $\vec v_1$ and
$\vec v_2$, and semi-axes $\lambda_1$ and $\lambda_2$. The geometrical
meaning of this ellipse is the following: a particle moving away from
the point at unit speed in the manifold ${\cal M}$ would move in
$\R^2$ with a speed given by the intersection of the ellipse with the
ray which the particle follows.

\begin{figure}
\centerline{\epsfig{file=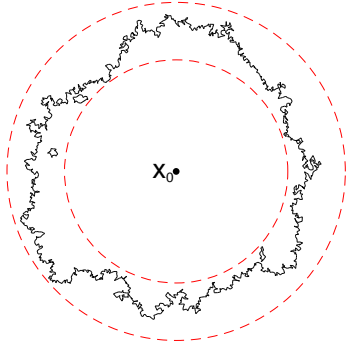,width=4.5cm,angle=0}}
\caption{\label{fig.ball} A typical ball $B_{X_0}(r)$ of radius $r$
  around a given point $X_0$. In general, $B_{X_0}(r)$ need not be
  simply connected. Therefore, its boundary $\partial B_{X_0}(r)$
  (solid lines) contains different components. The boundary is
  enclosed between two circles (dashed lines), whose radii grow
  linearly with $r$ \cite{Lagatta_10,Lagatta_14}.}
\end{figure}

Now let us choose a point $X_0$ (e.g., the origin) and consider the
set of points, $B_{X_0}(r)=\{X\;|\;d_g(X,X_0)\leq r\}$, whose
$g$-distance to it is smaller than or equal to a certain $r$. Since
$X_0$ will remain fixed from the beginning, we will usually drop the
subindex. This ball need not be topologically equivalent to an
Euclidean ball, since it need not be simply connected. Therefore, its
boundary $\partial B_{X_0}(r)$ will consist of a certain number of
components, see figure \ref{fig.ball} for a pictorial image.

The results in \cite{Lagatta_10,Lagatta_14} guarantee that when this
boundary, $\partial B_{X_0}(r)$, is viewed from the Euclidean
viewpoint, it lies within two circles centered at $X_0$, whose radii
scale linearly with $r$. It is not hard to prove that one of the
components of the ball boundary encloses all the others, namely, the
one whose interior contains $X_0$. Thus, the ball-boundary consists of
an outer irregular front plus an internal {\em froth}, or set of
bubbles. Let $\partial_0 B_{X_0}(r)$ denote this exterior component.

A useful mental image of the ball is a swarm of particles emanating
from $X_0$, each one escaping from there with unit speed and following
a geodesic line. At time $t$, the set of visited points will be
$B_{X_0}(t)$. In this way, the bubbles can be considered as ``hills''
which are hard to climb. This picture can be made more precise in the
following way. Let us consider the tangent space at $X_0$, $T_{X_0}$,
and the set of (outward) unit vectors, $\vec u_\varphi$, parameterized
by some angle $\varphi$. Each $\vec u_\varphi$ determines a unique
geodesic curve, $\gamma_\varphi$. If each geodesic is traversed at unit
speed, then time is a natural (arc-length) parameter for this curve,
$\gamma_\varphi(t)$, with $\gamma_\varphi(0)=X_0$. We now state that
$\partial B(t) \subseteq \cup_\varphi \gamma_\varphi(t)$. The equality
does not generally hold, since many geodesics are non-minimizing
\cite{Burns}.

In fact, $\{\varphi,t\}$ constitute a ---possibly degenerate---
coordinate system on the manifold that generalizes polar
coordinates. It has a very interesting property: lines of constant
$\varphi$ and lines of constant $t$ are always $g$-orthogonal. Building
from this assertion, one can state a modified {\em Huygens principle}
for the propagation of the ball front. Given the front at a certain
time $\partial B(t)$, it is possible to obtain the front at $t+\delta
t$, $\partial B(t+\delta t)$ by allowing each point $X$ on it to move
along the local normal direction $X\to X+\delta t\cdot \vec n$. This
is, of course, in analogy to the original Huygens principle for the
propagation of light, or Hamilton-Jacobi equation in mechanics.

Let us start with an infinitesimal circle centered at $X_0$. Then the
ball for {\em time} $t$ fulfills simply the equation

\begin{equation}
\partial_t X=\vec n_g(X),
\label{propagation}
\end{equation}
where $X$ stands for a generic point on $\partial B(t)$ and $\vec
n_g(X)$ is the local normal to such an interface, with respect to the
metric $g$.

\begin{figure}
\centerline{\epsfig{file=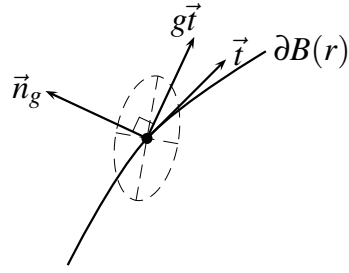,width=4.5cm,angle=0}}
\caption{\label{fig.tn} A small region of the ball front, showing the
  local tangent ($\vec{t}\/$) and normal ($\vec{n}_g$) vectors. They are
  orthogonal only with respect to the metric $g$, which is represented
  with the dashed ellipse. Notice that $\vec t$ and $\vec n_g$ are
  {\em not} orthogonal within the Euclidean framework. Instead, in the
  Euclidean metric $\vec n_g$ is orthogonal to $g\vec t$, which is the
  correct notion of $g$-orthogonality.}
\end{figure}

We can gain some intuition about this Huygens principle from figure
\ref{fig.tn}, which shows a zoom on a region of the ball front. The
dashed ellipse shows the local metric tensor $g$. How to obtain the
normal vector $\vec n_g$, given the tangent $\vec t$ and the metric?
The $g$-orthogonality relation $\vec t \perp_g \vec n_g$ can be stated
as $g_{\mu\nu} t^\mu n_g^\nu=0$, i.e., $g \vec t \perp \vec n_g$,
where $\perp$ denotes the Euclidean orthogonality
relation. Application of $g$ to $\vec t$ makes it always closer to the
principal direction with maximal eigenvalue. Therefore, $\vec n_g$
will always be closer to the principal direction with {\em minimal}
eigenvalue. Of course, $\vec n_g$ must be $g$-normalized, so that the
front will move with unit speed in ${\cal M}$.

Therefore, if the metric is given, the propagation algorithm can be
summarized as follows:

\begin{itemize}
\item{} For each point of the front, find $\vec t$.
\item{} Compute $g\vec t$.
\item{} Find an Euclidean normal to that vector, $\vec N_g$. Of course,
  take good care of the orientation!
\item{} Normalize that vector according to $g$, namely, find $|\vec
  N|^2_g\equiv g_{\mu\nu}N_g^\mu N_g^\nu$ and compute $\vec n_g=(1/|N|_g) \vec
  N_g$.
\item{} Move the point by the vector quantity $\delta t\cdot \vec n_g$.
\end{itemize}


\section{Numerical simulation algorithm}
\label{numerics}

We have adapted our intrinsic-geometry algorithm for the covariant KPZ
equation, employed in \cite{LSC_JSTAT11,SLC_PRE14}, to the simulation
of the balls in generic metrics. In our approach, we simulate the ball
propagation of equation (\ref{propagation}), starting out with an
infinitesimal circle, and allowing time to play the role of the ball
radius. The ball at any time will be given by a list of points on the
plane. The spatial resolution of the front is held constant: the
Euclidean distance between two neighboring points $\Delta x$ must stay
within a certain interval $[l_0,l_1]$. This is done by inserting or
removing points in a dynamical way. Checks of our results for
invariance under changes in $l_0$ and $l_1$ are performed in order to
guarantee that the continuum limit has been achieved. Moreover,
self-intersections can appear naturally, as anticipated in figure
\ref{fig.ball}. In such cases, we retain only the component which
contains the origin of the ball, i.e., we track $\pl_0 B(r)$.

\begin{figure}
\centerline{\epsfig{file=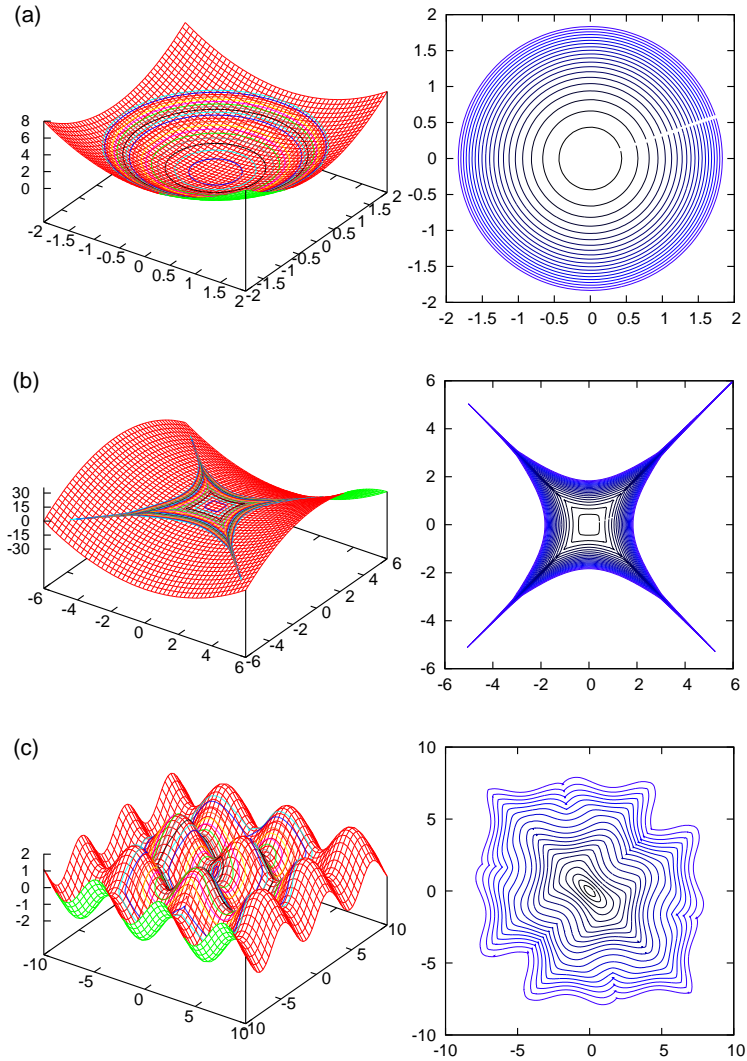,width=14cm,angle=270}}
\caption{\label{fig.detballs} Examples of use of our numerical
  algorithm to obtain balls with increasing sizes for deterministic
  metrics. In all cases, and in order to enhance visualization, the
  metrics have been extracted from the first fundamental form of a
  simple surface: (a) Paraboloid $z=x^2+y^2$, a surface with positive
  curvature; (b) saddle $z=x^2-y^2$, a surface with negative
  curvature; (c) egg-crate surface $z(x,y)=\sin(x)+\sin(y)$. Left
  column shows the balls (solid lines) immersed in the corresponding
  surfaces. The right column does the same in the Euclidean $(x,y)$
  plane.}
\end{figure}

For illustration, figure \ref{fig.detballs} shows the integration
procedure as applied to several deterministic metrics. In each case,
the metric $g$ is obtained from the first fundamental form of a simple
surface. Indeed, the form of the corresponding balls in the Euclidean
plane intuitively reflect the ``speed'' with which the interface
(ball) grows at each point as a function of the value of the metric
there.

Generation of a random metric tensor field is performed by assuming
that the correlation length is shorter than the cutoff distance
assumed for the ball, i.e., $r_0<l_0$. Thus, the metric tensors at
sampled points are statistically {\em independent}. The procedure does
not require derivatives of the metric tensors, as it would if one
insisted on tracking individual geodesics. The metric tensor at each
point is specified by providing the two orthogonal unitary
eigenvectors, $\vec v_1$ and $\vec v_2$, and the two corresponding
eigenvalues. Thus, $\vec v_1$ is generated randomly, $\vec v_2$ is
just chosen to be orthogonal to it, and the eigenvalues are
uniform deviates in the interval $[\lambda_0,\lambda_1]$, where
$\lambda_0$ should be strictly larger than zero.

The balls are analyzed from the Euclidean point of view: Their
roughness $W$ is found after fitting to an Euclidean circle, and by
computing the average squared deviation from the ball points to it,
ultimately averaging over disorder realizations. We will also consider
the standard deviation $\sigma_r$ of the radius of the fitting
Euclidean circle over realizations of the disorder
\cite{SLC_PRE14}. Thus, $W$ can be interpreted to quantify {\em
  intra}-sample radial fluctuations, while $\sigma_r$ assesses {\em
  inter}-sample radial fluctuations.


\section{Balls and geodesics in random metrics}
\label{balls}

The algorithm described in the previous section has been applied to
integrate equation (\ref{propagation}) numerically for different
realizations of the random geometry. Our simulations start with a very
small ball, with initial radius $0.05$, and propagate it through a
random metric with eigenvalues $\lambda \in [1/20,1]$. The time-step
used is $\Delta t=5\cdot 10^{-3}$ and the ultraviolet cutoff interval
for the simulation is chosen to be $[l_0,l_1]=[0.01,0.05]$. Scaling
results were checked to remain unchanged for smaller values of the
discretization parameters.

\begin{figure}[h]
\centerline{\epsfig{file=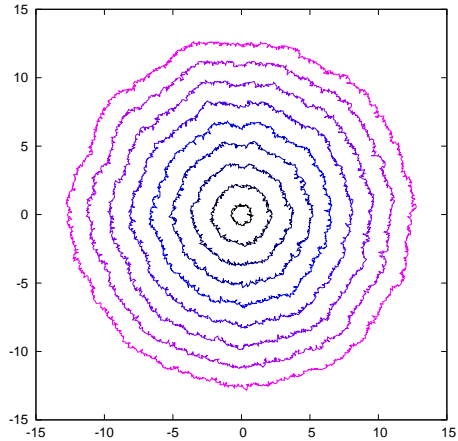,width=6cm,angle=270}}
\caption{\label{fig.profiles}Example of balls of increasing radii of a
  metric chosen randomly with eigenvalues $\lambda\in[1/20,1]$.
  The {\em times} or $g$-radii grow linearly from $t=0.2$ to $3.4$.}
\end{figure}

Figure \ref{fig.profiles} shows an example of balls with increasing
radii for times (i.e., $g$-radii) in the range $t=0.2$ to $3.4$.

\subsection{Roughness}

\begin{figure}[h]
\centerline{\epsfig{file=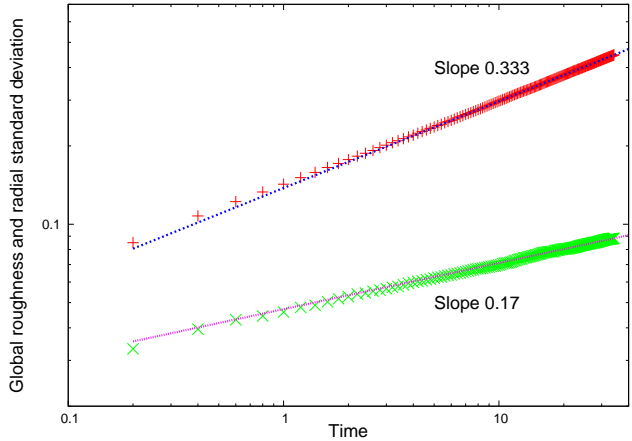,width=6cm,angle=270}}
\caption{\label{fig.rm.roughness} Log-log plot of Euclidean radial
  roughness $W$ ($+$) and standard deviation $\sigma_r$ ($\times$) as
  functions time (i.e., $g$-radius). Lines correspond to simple
  power-law behaviors, with exponent values as given in the
  corresponding labels.}
\end{figure}

We have simulated equation (\ref{propagation}) for 1280 realizations
of the disorder in the metric and analyzed the Euclidean roughness of
the resulting balls as a function of time (i.e., $g$-radius). The
results for the roughness $W(t)$ are shown in figure
\ref{fig.rm.roughness}. The figure also shows the time evolution of
the standard deviation of the average fitting Euclidean radius,
$\sigma_r$. Both observables are seen to follow power-law
behavior. Thus, $W\sim t^\chi$, with $\chi \simeq 1/3$ ($0.333\pm
0.001$). The correspondence between the $\chi$ exponent of random
geometry and the $\beta=1/3$ value which characterizes the KPZ
universality class is evident.

With respect to the sample-to-sample standard deviation of the average
Euclidean radii, $\sigma_r$, we also obtain a clear power-law, as
illustrated in fig.\ \ref{fig.rm.roughness}, but with a different
exponent value, namely, $\sigma_r\sim t^{\tilde\chi}$, with
$\tilde\chi = 0.17\pm 0.01$. An heuristic argument shows that this
value is actually also compatible with KPZ scaling. Indeed, let us
assume (1) that the radius grows linearly in time and (2) that the
correlation length scales as $\ell \sim t^{1/z}$, with $z=3/2$, as
expected within KPZ universality. Then, the number of {\em independent
  patches} on a single droplet will scale as $n_P \sim r/\ell \sim
t/t^{1/z} = t^{1/3}$. The sample-to-sample standard deviation should
then scale as the local fluctuations divided by the square root of the
number of patches, $W/\sqrt{n_P} \sim t^{1/3} / t^{1/6} =
t^{1/6}$. Our value for $\tilde\chi$ is compatible with this
prediction.

\subsection{Geodesic fluctuations}

\begin{figure}[h]
\centerline{\epsfig{file=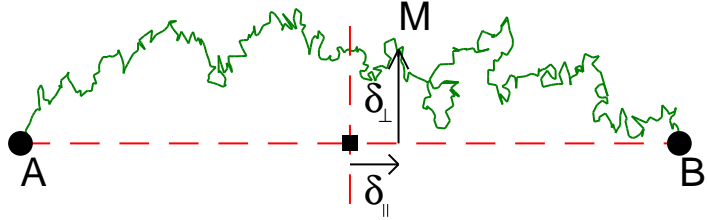,width=10cm,angle=0}}
\caption{\label{cartoon} Minimizing geodesic (solid line) between
  points $A$ and $B$, which are an Euclidean distance $L$ apart. Let
  $M$ be the middle point in the geodesic, i.e., the point along it
  which can be reached in the same time from $A$ and from $B$. Then,
  $\delta_\perp$ and $\delta_\parallel$ are the lateral and
  longitudinal deviations between $M$ and the middle point (square) of
  the straight segment (dashed horizontal line) joining $A$ and
  $B$. In other terms, they constitute the lateral and longitudinal
  deviations of the actual geodesic from its Euclidean counterpart.}
\end{figure}

Our next numerical experiment addresses the average lateral deviation
of the minimizing geodesics. We define such a geodesic fluctuation in
the following way. Consider two points which are an Euclidean distance
$L$ apart. Find the minimizing geodesic joining them, and mark also
its middle point $M$, see figure \ref{cartoon}.

The coordinates of point $M$ relative to the middle point of the
straight segment joining $A$ and $B$, namely
$(\delta_\parallel,\delta_\perp)$, are, respectively, the longitudinal
and lateral fluctuations of the actual geodesic from its Euclidean
counterpart. Notice that the disorder averages of both
$\delta_\parallel$ and $\delta_\perp$ should be zero, but their
fluctuations are highly informative. According to previous work
\cite{Lagatta_10,Lagatta_14}, they are conjectured to scale as
$\delta_\parallel\sim L^\chi$, with the same exponent as the
roughness, $\chi=1/3$, and $\delta_\perp \sim L^\xi$, with $\xi=2/3$,
a second critical exponent.

\begin{figure}[h]
\centerline{\epsfig{file=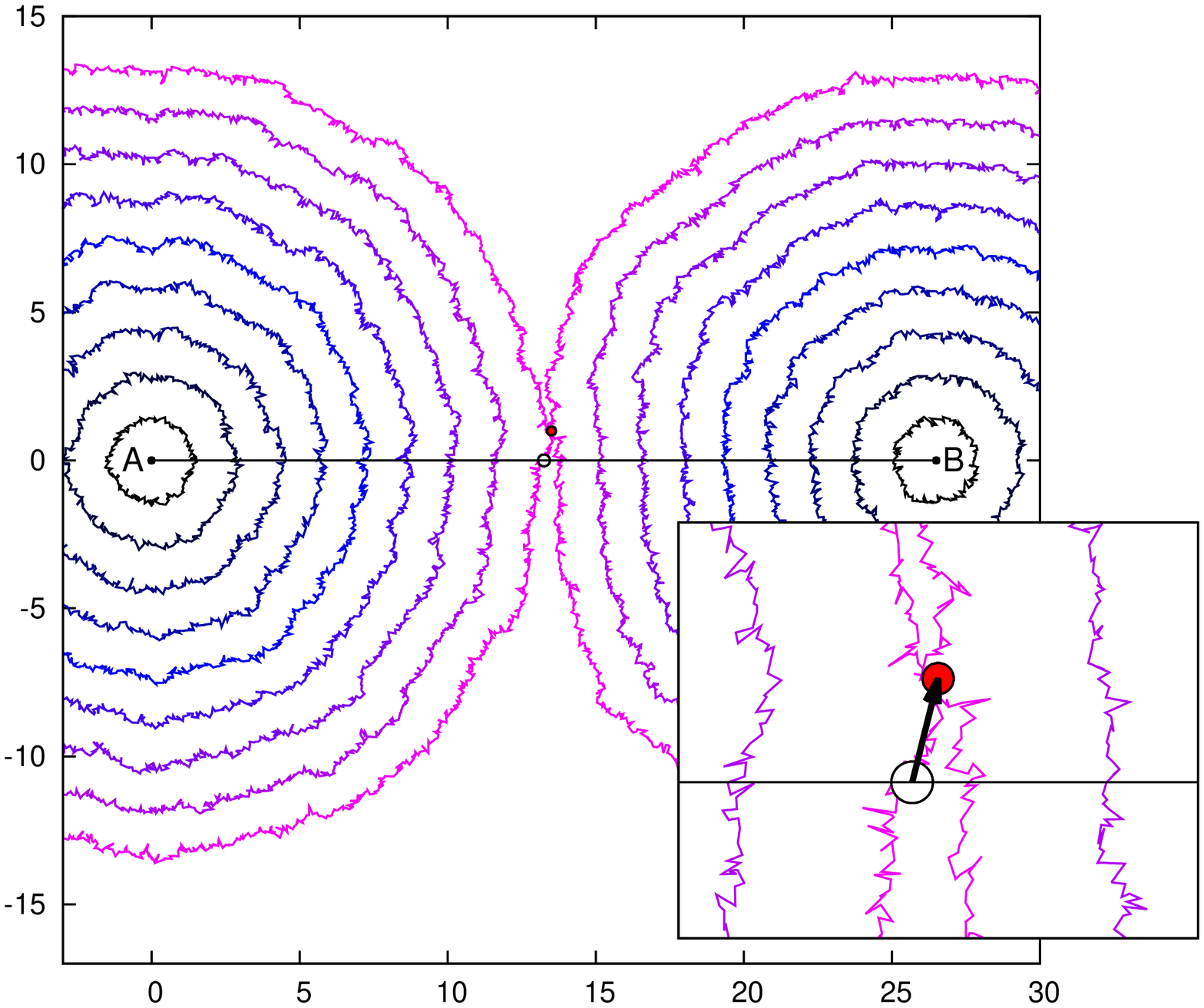,width=6cm,angle=270}
\epsfig{file=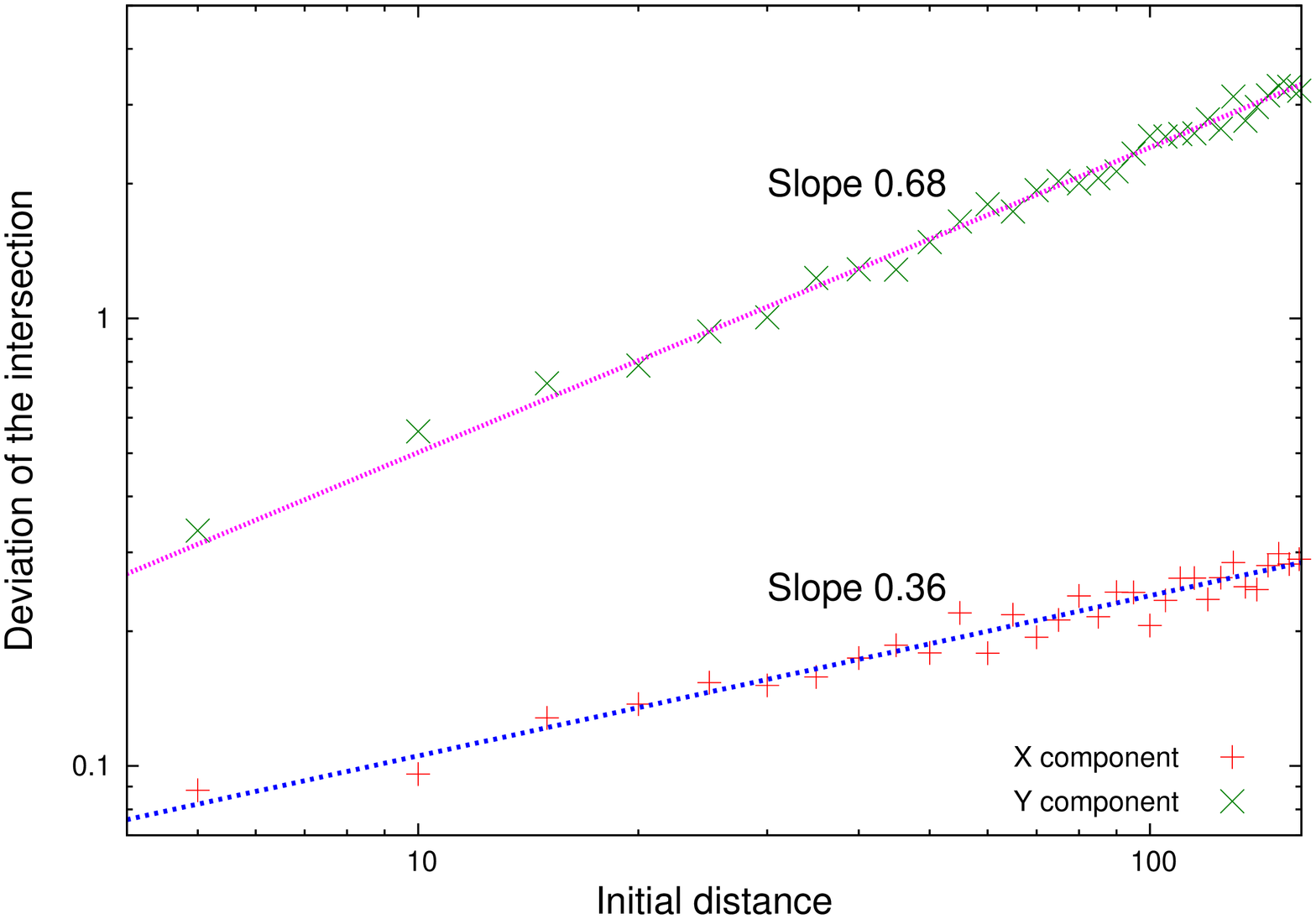,width=6cm,angle=270}}
\caption{\label{fig.twoballs}Left: Illustration of the procedure to
  find the geodesic fluctuations. Balls are grown simultaneously from
  points $A$ and $B$. Their first intersection point is marked (solid
  bullet). The vector going from the midpoint on the segment that
  joins the two ball centers (empty circle) to the intersection point
  has components $\delta_\parallel$ and $\delta_\perp$, see inset for
  a zoomed image. Right: Root-mean-square deviation of the
  intersection of the balls growing from two points at a distance $L$,
  in logarithmic scale, for $128$ samples. The lateral fluctuation
  ($\times$) scales as $\delta_{\perp} \sim L^\xi$, with $\xi=0.68\pm
  0.02$. The longitudinal fluctuation ($+$) scales as
  $\delta_{\parallel} \sim L^\chi$, with a smaller exponent
  $\chi=0.36\pm 0.02$. The lines provide power-law fits using these
  exponent values.}
\end{figure}

We have estimated both $\delta_\perp$ and $\delta_\parallel$, fixing
points $A, B$ at $(\mp L/2,0)$, for different values of $L$ and 128
realizations of the disorder. Our procedure is as follows: Two balls
centered at these points are grown simultaneously. Growth is arrested
when both balls intersect for the first time. The coordinates of their
first intersection point are, precisely,
$M=(\delta_\parallel,\delta_\perp)$; see figure \ref{fig.twoballs}
(left) for an illustration. The rationale is as follows. Let us call
$t_x$ the time ($g$-radius) at which both balls first intersect. Point
$M$ can be reached in time $t_x$ both from $A$ and from $B$, hence it
should belong to the minimizing geodesic connecting both points.

In order to save simulation time, each simulation is carried out in
practice as follows: We start with two very small balls separated by a
small distance $L_0$, and grow them until they first intersect. At
this moment, we take note of the coordinates of the intersection
point, increase the separation of the balls by $\Delta L$, rotate each
one by a random angle, and continue the simulation until they
intersect again. This procedure is repeated until the desired range
for $L$ has been covered. The random rotation ensures that the ensuing
intersection points are uncorrelated.

We obtain the root-mean-square horizontal and vertical deviations of
the intersection point as functions of the Euclidean distance between
the two points. The results appear in figure \ref{fig.twoballs}. The
lateral fluctuations of the geodesics scale with the separation $L$
between the ball centers as $\delta_\perp\sim L^\xi$, with $\xi\simeq
2/3$, while the longitudinal fluctuations scale with the same exponent
value as the roughness, namely, $\delta_\parallel\sim L^\chi$, with
$\chi\simeq 1/3$. It is straightforward to understand the exponent for
the longitudinal fluctuations, as $\delta_\parallel$ is quite
naturally expected to grow with the ball roughness. Through the rough
interface interpretation mentioned above \cite{Krug_92}, the lateral
fluctuations $\delta_\perp$, are otherwise related to the increase in
the {\em correlation length} characteristic of systems in the KPZ
universality class, namely, $\delta_\perp(t) \sim \ell(t) \sim
t^{1/z}$, with $1/z=2/3$. Indeed, the exponent values we obtain for
the random metrics system are compatible, within statistical
uncertainties, with the so-called Galilean relation, $\chi+1=2\xi$ or,
equivalently, $\beta + 1 = 2/z$, which is a hallmark of the KPZ
universality class.  The geometrical interpretation of this exponent
identity within the latter context is the expression, under the
scaling hypothesis, of the fact that on average the rough interface
grows with uniform speed along the local normal direction
\cite{Krug_92}, implementing a Huygens principle as discussed above.

\subsection{Rightmost point statistics}

\begin{figure}[h]
\centerline{\epsfig{file=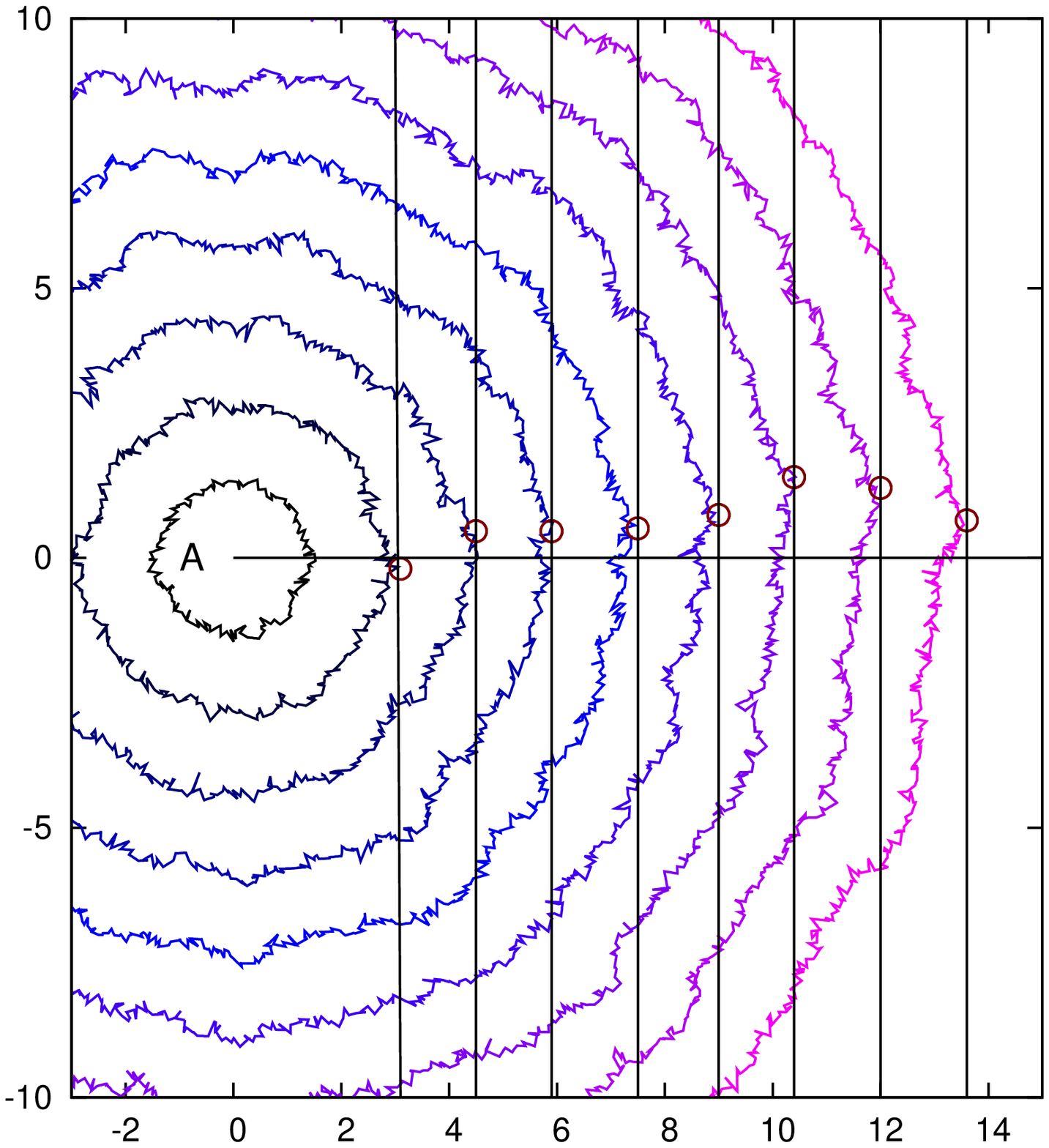,width=6cm,angle=270}
\epsfig{file=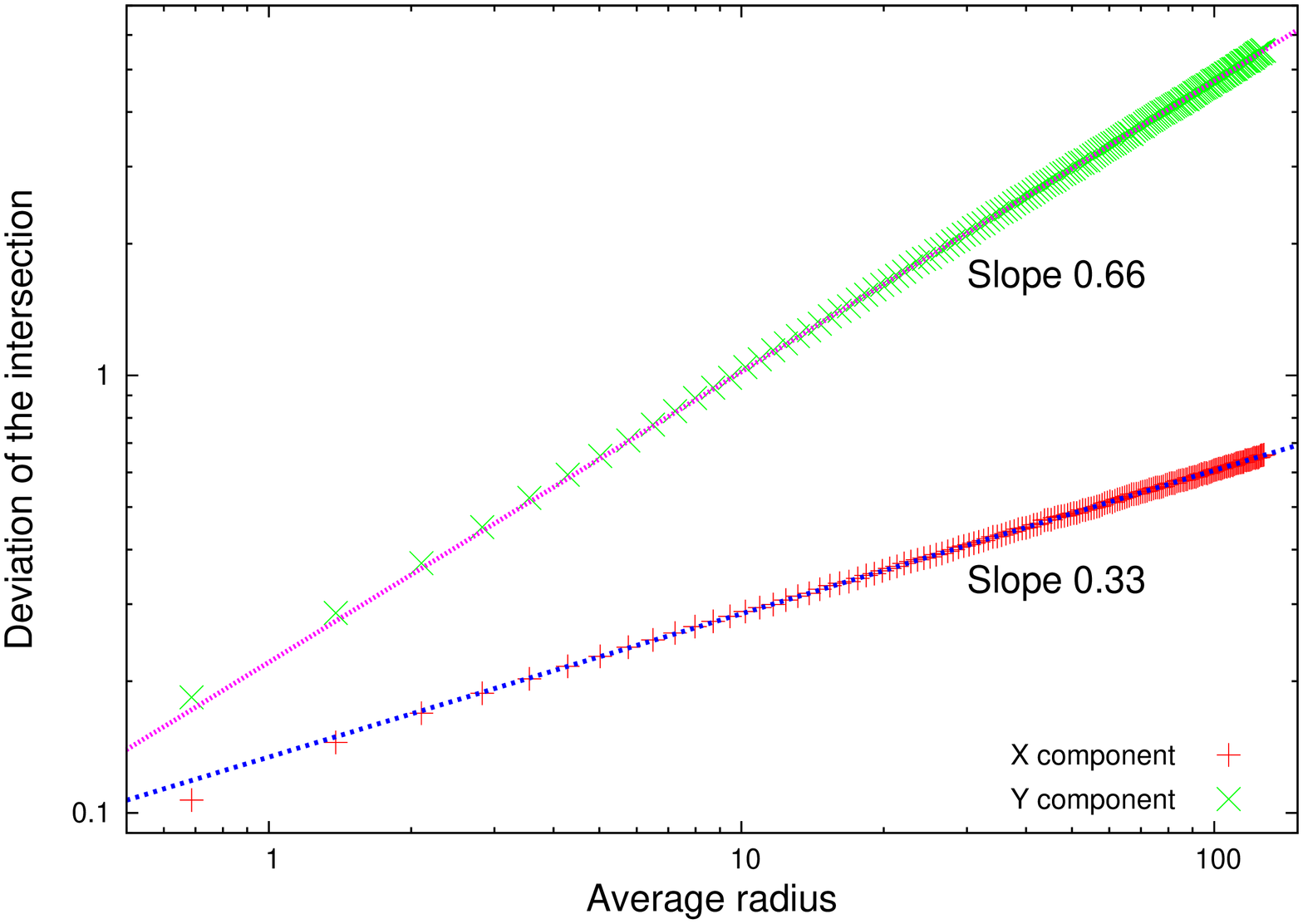,width=6cm,angle=270}}
\centerline{\epsfig{file=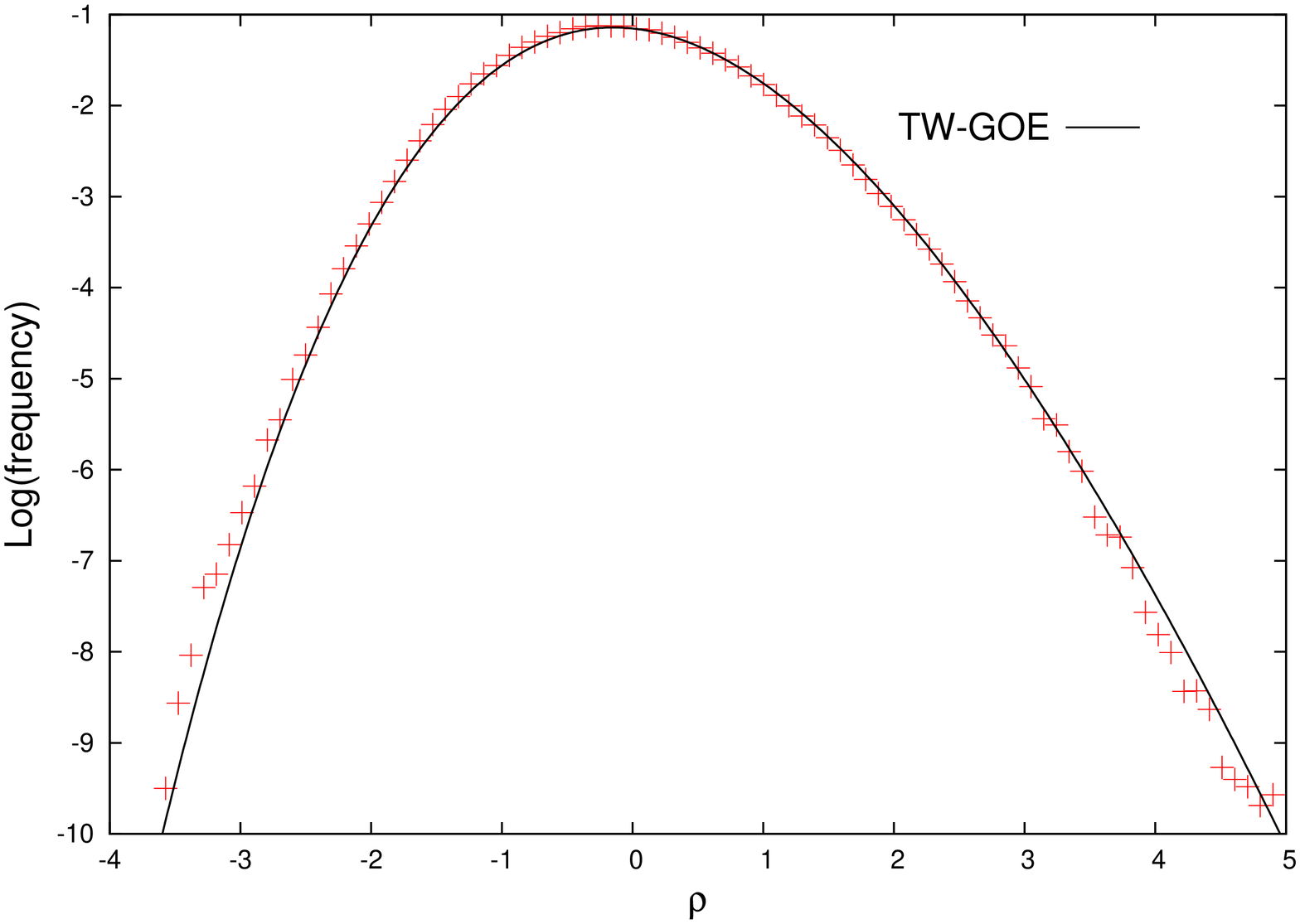,width=6cm,angle=270}}
\caption{\label{fig.rightmost} Upper left: Illustration of the
  rightmost-point procedure to estimate the transverse and
  longitudinal fluctuations of our interfaces. For each curve, the
  open symbol corresponds to its right-most point. Upper right: Scaling of
  the fluctuations with the average radius for each time, using 1280
  noise realizations and 50 different rotations per sample. Bottom:
  full probability distribution for the (rescaled) horizontal position
  of the rightmost-point, $\delta_\parallel$, to have zero mean and
  variance one. Comparison is provided to the TW-GOE distribution with
  the same normalization.}
\end{figure}

We have experimented with a different approach to find the scaling
exponents. For each interface, we find the {\em rightmost point},
$\vec P$, to be the point with highest $x$-coordinate. If the
interface was a circumference, we would have $\vec P=(R,0)$, where $R$
is the expected value of the radius. Let us write the deviations as
$\vec P=(R+\rho_\parallel,\rho_\perp)$, with $\rho_\parallel$ and
$\rho_\perp$ having similar interpretations to those discussed for
$\delta_\parallel$ and $\delta_\perp$ in the previous section.  Of
course, there is nothing special with the $x$-axis, one may choose any
direction. A useful strategy is to perform several random rotations of
the interface and find the rightmost extreme point for each of them,
computing the root mean square values for $\rho_\parallel$ and
$\rho_\perp$. Figure \ref{fig.rightmost} shows the results of this
procedure as a function of the average radius size for 1280
realizations, with 50 random directions for each profile. The results
are extremely clean: $\rho_\parallel \sim R^\beta$ with $\beta = 0.333
\pm 0.001$ and $\rho_\perp \sim R^{1/z}$ with $1/z = {0.665\pm
  0.001}$, fully compatible with KPZ scaling if we consider that
$R\sim t$.

Furthermore, we have studied also the full probability distribution
for this $\delta_\parallel$, as shown in the bottom panel of
Fig.\ \ref{fig.rightmost}. The distribution follows a Tracy-Widom type
form, but not for the GUE as it is the case for circular fluctuations,
but for the GOE. Indeed, the numerical values for skewness
and kurtosis are 0.293 and 0.175, fully compatible with the TW-GOE
values (0.29346 and 0.16524). This result is analogous to those obtained
when considering the extreme-value statistics of the height of {\em curved} interfaces in e.g.\ experiments on turbulent liquid crystals \cite{Takeuchi.JSP.12} or in the Polynuclear Growth Model \cite{Johansson.CMP.03}, see additional references in \cite{Takeuchi.JSP.12}.


\section{Radial fluctuations}
\label{airy}

As discussed in the introduction, physical systems for which
fluctuations belong to the 1D KPZ universality class are consistenly
being found to not only share the values of the critical exponents
$\beta$ and $1/z$ (respectively, $\chi$ and $\xi$ in the random metrics
language), but also to be endowed with a larger universality trait,
alike to a central limit theorem: Radial fluctuations of the interface
follow the same probability distribution function as the largest
eigenvalues of large random matrices extracted from the Gaussian
unitary ensemble (GUE), i.e., the Tracy-Widom GUE (TW-GUE)
distribution. Moreover, higher-order correlations are also conjectured
to be part of the universality class, constituting the so-called Airy-2
process \cite{Praehofer_02,Corwin_13}.

\begin{figure}[h]
\centerline{\epsfig{file=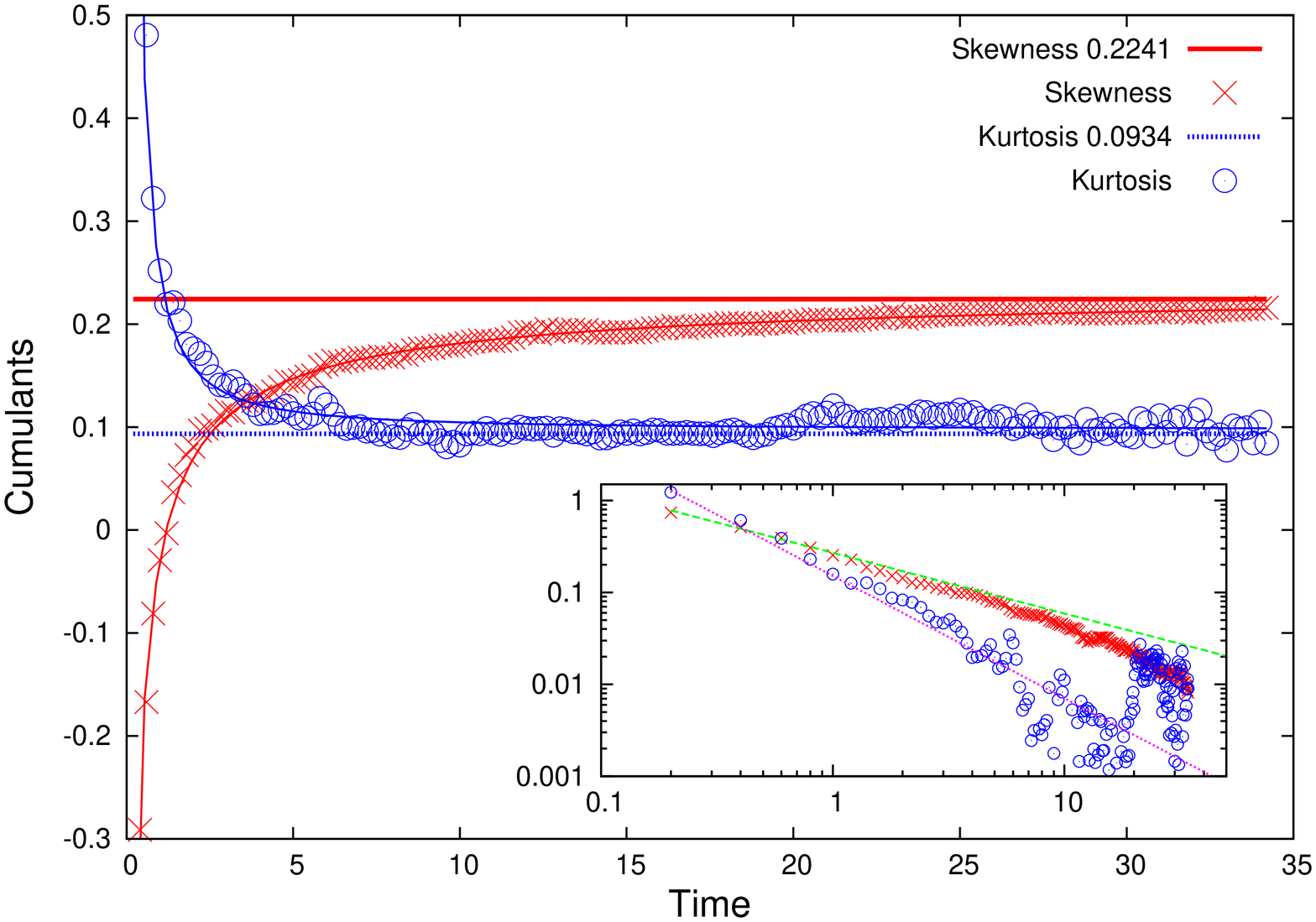,width=6cm,angle=270}
\epsfig{file=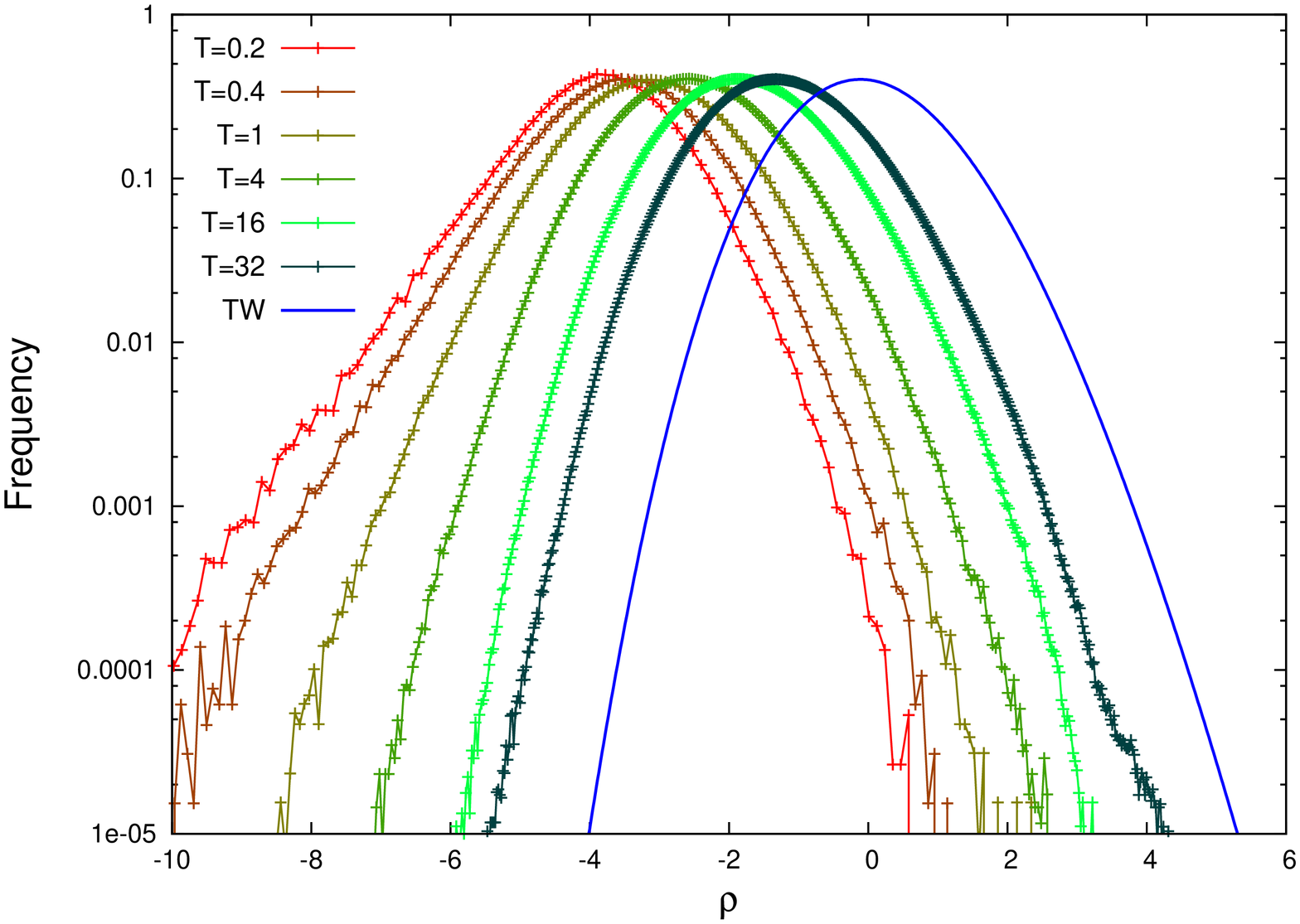,width=6cm,angle=270}}
\caption{\label{fig.radial_fluct}Left: Cumulants [skewness ($\times$)
    and excess kurtosis ($\circ$)] of the distribution of fluctuations
  in the Euclidean distances to the origin of points of balls as
  functions of time (i.e., $g$-radius). As references, the horizontal
  lines provide the asymptotic values corresponding to the TW-GUE
  distribution. Convergence is in the form of power-laws (solid
  curves), approximately $t^{-2/3}$ for the skewness and $t^{-4/3}$
  for the kurtosis. The inset is a log-log representation of the same data,
  in which the latter rates are shown as the green dashed and pink dotted lines,
  respectively. Right: Full histogram of the rescaled fluctuations
  in Euclidean distances to the origin, $\rho_i$, equation
  (\ref{def_rho}), for times as in the legend (symbols). Consistent
  with the evolution of their cumulants, the distributions are seen to
  approach the TW-GUE pdf (solid line).}
\end{figure}

Figure \ref{fig.radial_fluct} characterizes the probability
distribution function of the fluctuations in the Euclidean distances
to the origin of points on balls, for different times (or
$g$-radii). We consider 171 different values of time $t_i$, separated
by $0.2$ units, up to $t_{max}=35$ and discarding the first few
times. For each time, the corresponding ball is approximated by a
circle with radius $r(t)$, which in turn is fit to the deterministic
shape $r(t)=r_0+vt$. An average over 1280 noise realizations is
made. Once such a deterministic contribution to radial growth is
identified, we subtract it from the distance to the origin $r_i$ of
each point on the corresponding ball, as

\begin{equation}
\rho_i \equiv {r_i - r_0 - v t \over \Gamma t^\gamma},
\label{def_rho}
\end{equation}
where $\Gamma$ is a normalization constant. The Pr\"ahofer-Spohn
conjecture \cite{Praehofer_PRL00,Praehofer_PA00} for the radial
fluctuations $\rho_i$ within the KPZ universality class is that
$\gamma$ should be equal to $\beta=1/3$, with a pdf which converges
for long times to the TW-GUE distribution. Figure
\ref{fig.radial_fluct} (left) shows the evolution of the third and
fourth cumulants of the distribution of radial fluctuations. They both
can be seen to approach asymptotically their TW-GUE values, which are,
respectively, $0.224$ and $0.093$
\cite{Praehofer_PRL00,Praehofer_PA00}.  Within our statistics, the
decay rates for the differences between our estimates for skewness and
kurtosis and their asymptotic values are compatible with power-law
rates $t^{-2/3}$ and $t^{-4/3}$, respectively, see the inset of the
left panel in figure \ref{fig.radial_fluct}. For the skewness, this
convergence rate seems to agree with results in e.g.\ some discrete
growth models \cite{Ferrari_JSP11} and in experiments
\cite{Takeuchi.JSP.12}, while this is not the case for the
kurtosis. Although further studies are needed, these finite time
corrections to the distribution seem likely not to be universal; see
additional results e.g.\ in \cite{Alves_JSTAT13}. Considering the full
histogram of the fluctuations, its time evolution is shown on the
right panel of figure \ref{fig.radial_fluct}, wherein steady
convergence to the TW-GUE distribution can be readily appreciated.

\begin{figure}
\centerline{\epsfig{file=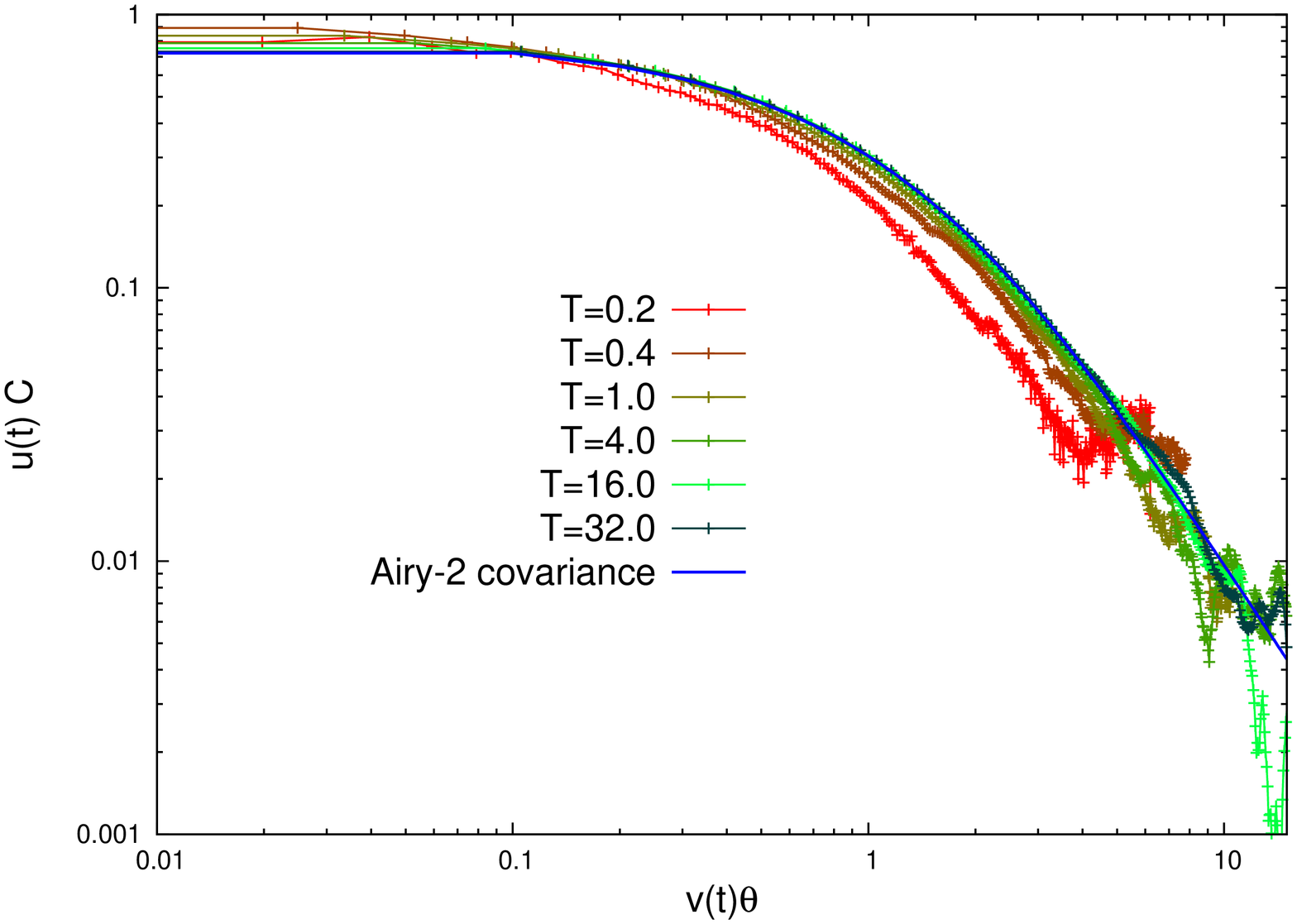,width=6cm,angle=270}
\epsfig{file=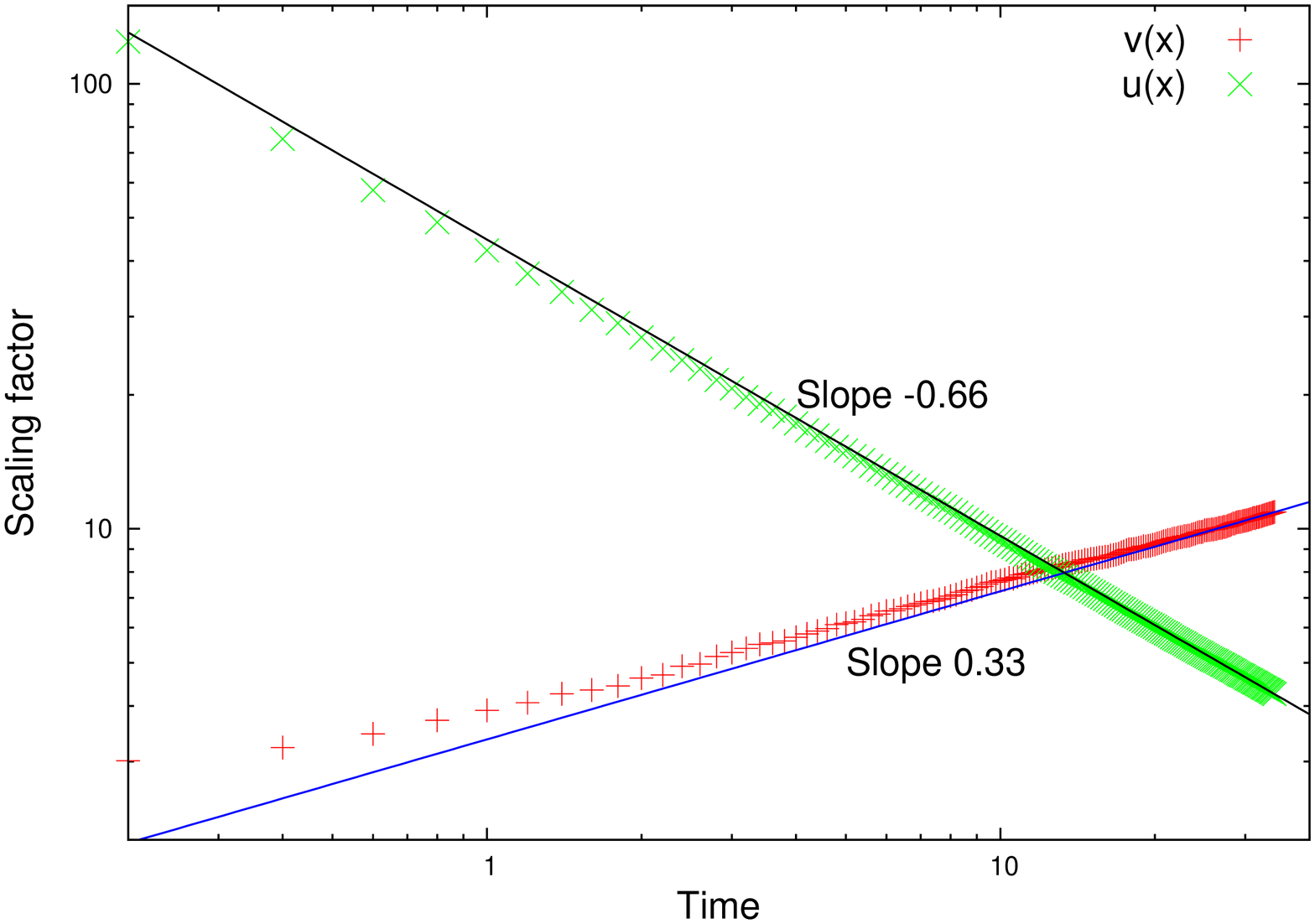,width=6cm,angle=270}}
\caption{\label{fig.correlator}Left: Data collapse of the angular
  correlation functions $C(\theta,t)$ for different times, to the
  Airy-2 covariance $g_2$, as in eq.\ \ref{fit.corrairy}. Right:
  Time evolution of the $u(t)$ and $v(t)$ factors employed for the
  collapse in the left panel, showing power-law behavior as described
  in eq. \ref{fit.parameters}, namely, $u(x)\sim t^{-2/3}$ and $v(t)\sim
  t^{1/3}$.}
\end{figure}

But the Airy-2 process involves more than the single-point
fluctuations. In particular, the angular two-point correlation
function can also be predicted. For a fixed time $t$, let
$r(\theta,t)$ be the maximal radius at a given polar angle $\theta$
for the ball corresponding to $g$-radius equal to $t$. Then, we define
the correlator as

\begin{equation}
C(\theta,t) \equiv \< r(\theta_0,t) r(\theta_0+\theta,t) \> - \<r\>^2.
\label{correlator}
\end{equation}
In our simulations, see figure \ref{fig.correlator}, we have found
that the two-point functions obtained for different times collapse
into an universal curve through

\begin{equation}
u(t) C(\theta,t) \approx g_2( v(t) \theta),
\label{fit.corrairy}
\end{equation}
where the scaling function, $g_2$, is the covariance of the Airy-2
process
\cite{Alves_EPL11,Oliveira_12,Bornemann_MC10,Nicoli_13}. The $u(t)$
and $v(t)$ factors have been found numerically through the collapse of the
two-point data, and their values are plotted on the right panel of figure \ref{fig.correlator}, showing that

\begin{equation}
u(t)=A_u t^{-2/3} , \qquad v(t)=A_v t^{1/3} ,
\label{fit.parameters}
\end{equation}
where $A_{u,v}$ are constants. The $v(t) \sim t^{1/3}$ dependence of the rescaled angular variable is related to the growth of the correlation length in a
straightforward way: a length variable would require a scaling factor
given by $t^{-1/z}=t^{-2/3}$. But since $\theta$ is an angle and the
radius scales linearly in time, the scaling factor is modified to
$t^{1-1/z}=t^{1/3}$. As seen in fig.\ \ref{fig.correlator},
convergence to the covariance of the Airy-2 process is indeed obtained
for sufficiently long times.

\section{Time of arrival}
\label{timeofarrival}

As mentioned in the introduction, first passage percolation
(FPP) systems bear a strong relation to the random metric problem
studied here \cite{Lagatta_10}. Indeed, the random passage times
between neighboring sites can be considered to constitute a
discretization of a Riemannian metric. The most important observable
in FPP studies is typically the {\em time of arrival} to different
sites in the lattice, which can be associated to the length of the
minimizing geodesic joining the origin to the given point
\cite{Kesten_03,Johansson_00,Chatterjee_13}.

\begin{figure}
\centerline{\epsfig{file=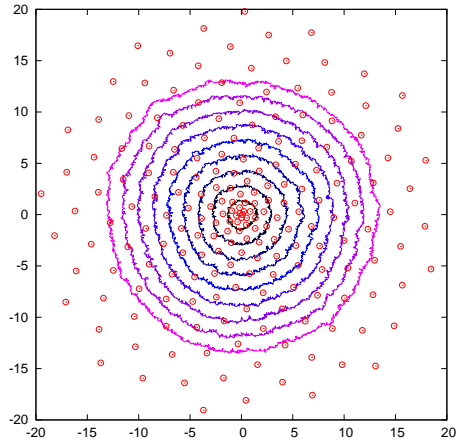,width=6cm,angle=270}}
\caption{\label{fig.toa_illust} Illustration of the checkpoint
  distribution employed in our algorithm to measure the time of arrival in the
  random metric system.}
\end{figure}

Measuring times of arrival within our scheme requires a special
simulation device, illustrated in figure \ref{fig.toa_illust}. A
number of {\em checkpoints} $X_j$ have been scattered throughout the
manifold. At each time step, a winding-number algorithm is performed
in order to check whether each one of them is {\em inside} or {\em
  outside} the corresponding ball. When point $X_j$ changes status
from outside to inside, we identify that time as its arrival time. The
points $X_j$ are distributed as a linear golden spiral, i.e., their
Euclidean radii increase linearly, but their angles follow the
sequence $\alpha_j = 2\pi j \phi $, where $\phi$ is the golden
section, $\phi=(\sqrt{5}-1)/2$. This distribution is chosen so as to
ensure a uniform angular distribution, as uncorrelated as possible.

\begin{figure}
\centerline{\epsfig{file=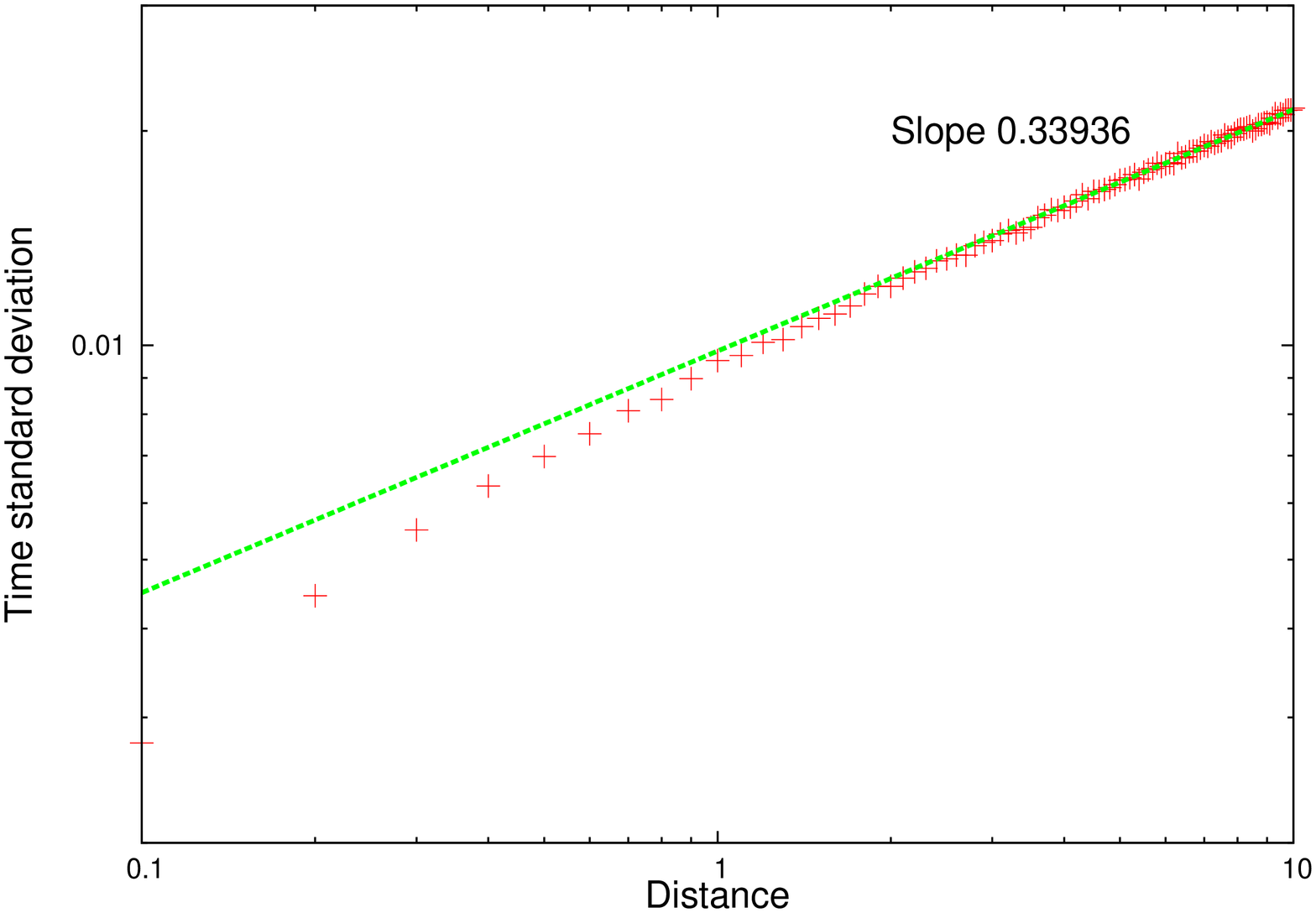,width=6cm,angle=270}
\epsfig{file=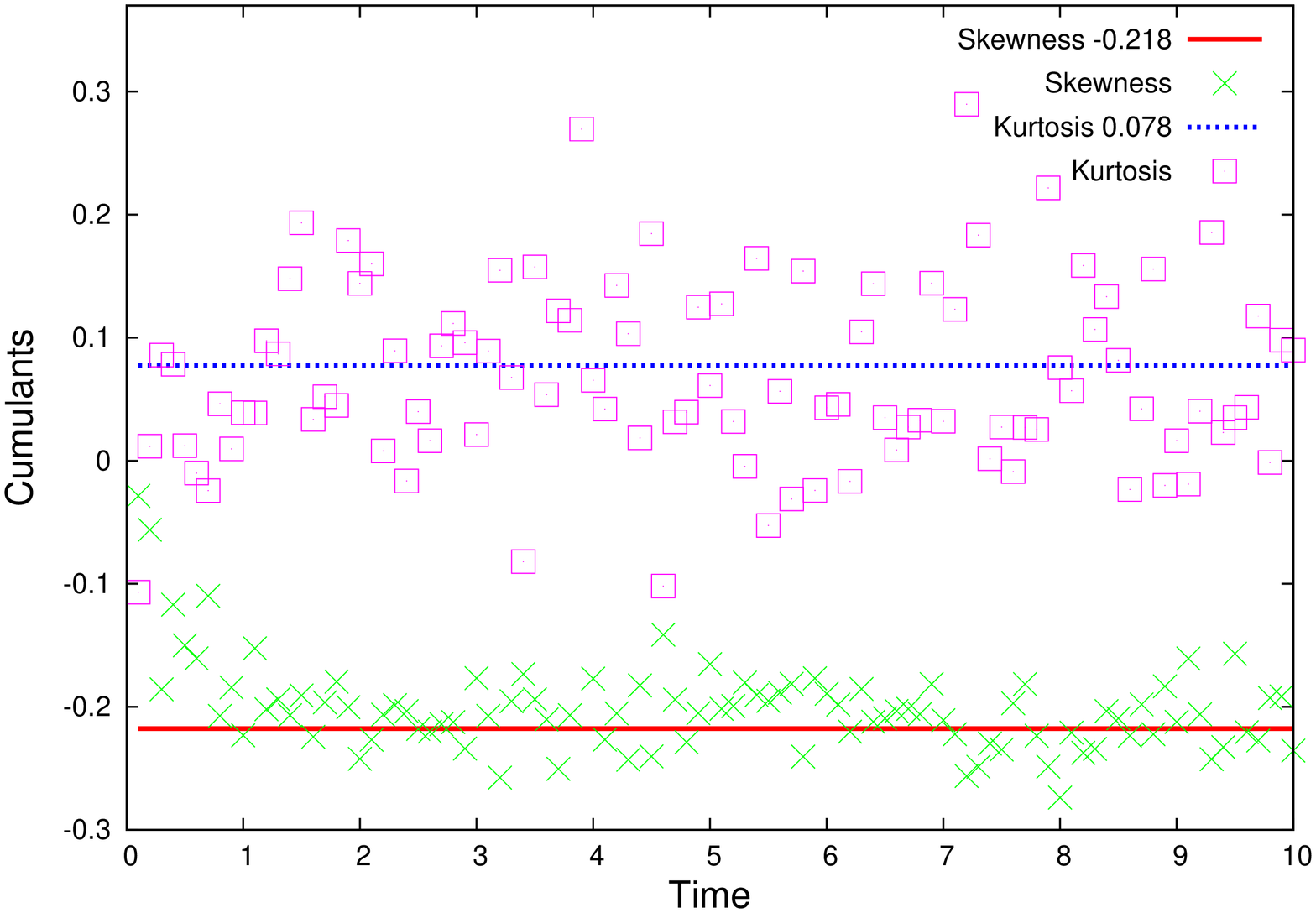,width=6cm,angle=270}}
\centerline{\epsfig{file=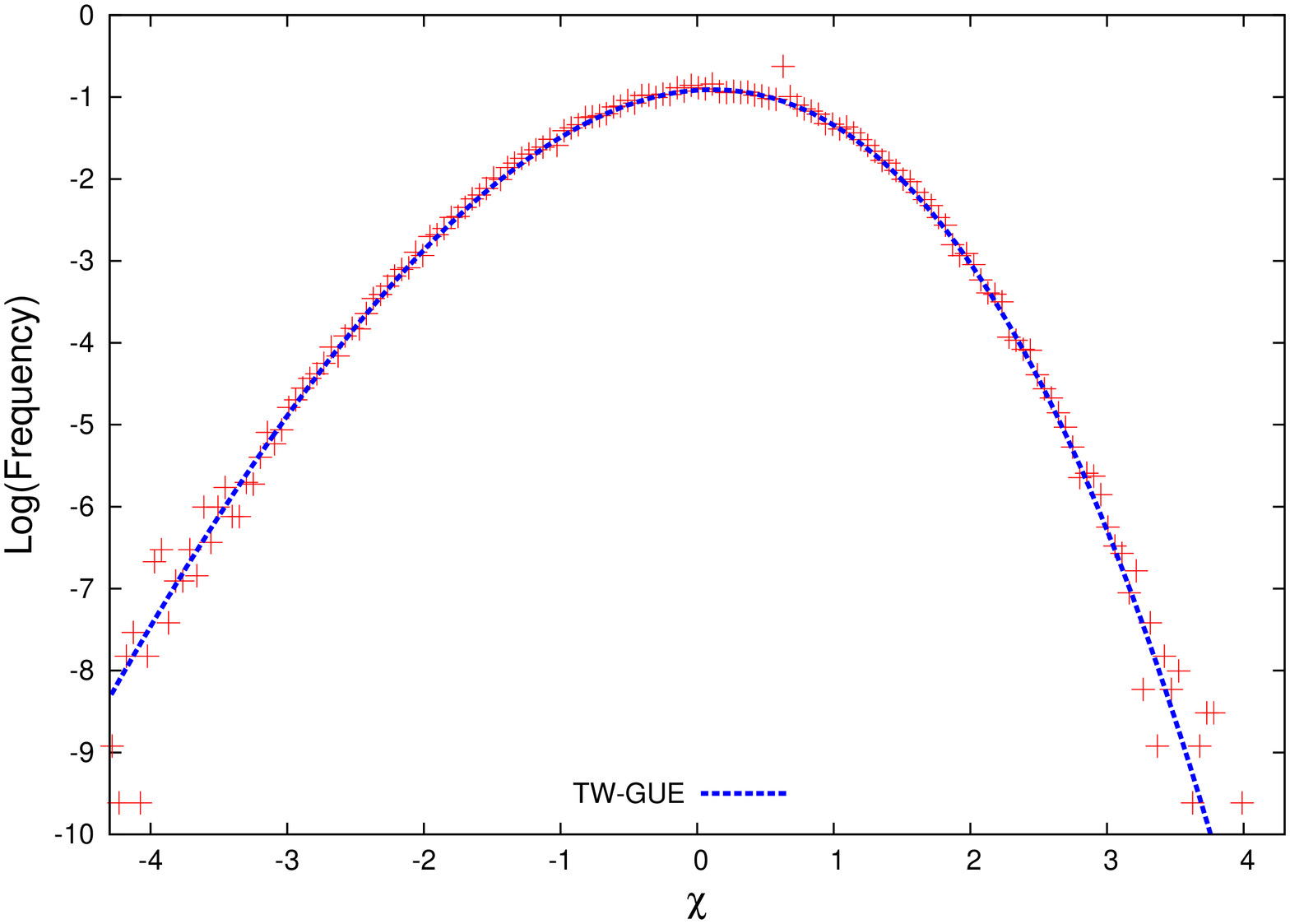,width=6cm,angle=270}}
\caption{\label{fig.toa} Numerical results for the distribution of
  times of arrival in the random metrics system. Top-left: Deviation
  of the time of arrival ($+$) as a function of Euclidean distance to
  the center of the ball. The slope of the straight line is provided
  in the legend. Top-right: Cumulants [skewness ($\times$) and kurtosis
    ($\Box$)] of the time distribution of arrival times as a
  function of the Euclidean radius. Exact TW-GUE values are given by
  the horizontal straight lines, for reference. Bottom: Histogram of
  the time of arrival fluctuations ($+$), and comparison with the
  TW-GUE distribution (dashed line).}
\end{figure}

The numerical simulations give the expected results, namely, the times
of arrival grow linearly with distance to the origin, and their
standard deviation also increases with distance, as $\sigma_t \sim
d^{0.339}$, see figure \ref{fig.toa} (upper left panel). The higher
cumulants are compatible with the TW-GUE distribution, averaging to
$-0.218$ for the skewness and 0.078 for the kurtosis, see upper right
panel. These values are to be compared with those obtained in
\cite{Praehofer_PRL00,Praehofer_PA00}, $-0.224$ and $0.093$,
respectively, as in the case of the radial fluctuations studied
earlier in fig.\ \ref{fig.radial_fluct}. The full histogram of arrival
times, and its comparison with the TW-GUE distribution, is shown in
figure \ref{fig.toa} (lower panel). Excellent agreement is obtained.

\section{Conclusions and further work}
\label{conclusions}

We have shown evidence of KPZ scaling in a purely geometric model, in
which the role of the evolving interface is played by balls of
increasing radii in a random manifold. Universal behavior occurs at
distances which are large as compared to either the correlation or the
curvature lengths. When the balls on the random manifold are viewed
from an Euclidean point of view, they appear to be rough. If the
radius of the balls is thought of as time, we show that the growth of
the Euclidean roughness of the ball is $W\sim t^\chi$, with
$\chi=1/3$. Moreover, study of the minimizing geodesics has shown that
the lateral correlations of the fluctuations in the balls scale as
$\ell \sim t^\xi$ with $\xi=2/3$. These critical exponent values are
the hallmark of the KPZ universality class, although in a different
language, namely, $\chi \to \beta$ and $\xi \to 1/z$. Our results thus
allow to assess numerically the predictions for Riemannian
first-passage percolation \cite{Lagatta_10,Lagatta_14}, providing a
detailed picture of its stochastic behavior. Given the relation to FPP
proper, this detailed description may aid in the development of
rigorous proofs that fully justify the values of the wandering and
fluctuation exponents in this important discrete model.

In principle, the results we obtain may come as a surprise. The random
geometry model we study features {\em quenched disorder}, i.e., the
disorder does not change with time. However, we obtain standard KPZ
universality (namely, the critical behavior corresponding to
time-dependent noise), which differs from the so-called quenched-KPZ
universality class \cite{Barabasi}. The latter describes the scaling
behavior of e.g.\ the quenched KPZ equation, in which a depinning
transition occurs: If the intensity of the external driving $F$ is
below a finite threshold $F_c$, then the average interface velocity
$v$ is zero. On the contrary, the interface moves with a non-zero $v$
for $F> F_c$. Actually, for sufficiently large $F$ the quenched
disorder is somehow seen by the interface as time-dependent noise, and
the scaling behavior becomes standard KPZ
\cite{Barabasi}. Quenched-KPZ scaling applies at the depinning
threshold $F=F_c$ \cite{Tang92_etal}. In the model we study the
average interface velocity is non-zero by construction, so that one is
always in the moving phase, in such a way that seemingly only standard
KPZ behavior ensues. Given the relation of the random geometry model
with FPP, and in turn the connection of the latter with the Eden
model, it is natural to ponder whether our results may provide some
clue on the relation between the quenched and the time-dependent KPZ
universality classes. Note that TW fluctuations have been also found in
other paradigmatic systems of quenched disorder, such as spin glasses,
structural glasses, or the Anderson model
\cite{Castellana,Somoza_07}. This point seems to warrant further
study.

Actually, TW statistics do appear in our model both in the Euclidean
fluctuations of the random balls and in the random-metric fluctuations
of the Euclidean balls, which can be described as fluctuations in the
arrival times at different distances from the origin. Again the
deviation of these values follows the same power-law as the roughness
in standard KPZ growth, namely, $\sigma_t \sim t^\chi$, with
$\chi=1/3$, and the fluctuations follow TW statistics. Due to the wide
connections and applications of the FPP model to disordered systems,
one can speculate whether this reinterpretation might allow to unveil
TW statistics in still many other phenomena in which it has not been
identified yet. This would strengthen the role of TW fluctuations as a
form of a central limit theorem for many far-from-equilibrium
phenomena.

The code used to carry out the simulations in this work has been
uploaded as free software to a public repository \cite{Git}.


\ack We want to acknowledge very useful discussions with K.\ Takeuchi
and S.\ Ferreira. This work has been supported by the Spanish
government (MINECO) through grant FIS2012-38866-C05-01. J.R.-L.\ also
acknowledges MINECO grants FIS2012-33642, TOQATA and ERC grant
QUAGATUA. T.L.'s research and travel was supported in part by NSF PIRE
grant OISE-07-30136.



\section*{References}

\begin{thebibliography}{10}

\bibitem{Adler} R. Adler and J. Taylor, {\em Random fields and
  geometry}, Springer (2007).

\bibitem{Itzykson_Drouffe} C. Itzykson and J.-M. Drouffe, {\em
  Statistical Field Theory}, Cambridge University Press (1991).

\bibitem{Booss.book} B. Boo\ss-Bavnbek, G. Esposito and M. Lesch, {\em
  New paths towards quantum gravity}, Springer (2009).

\bibitem{Nelson_et_al} D. Nelson, T. Piran and S. Weinberg, {\em
  Statistical Mechanics of Membranes and Surfaces} World Scientific,
  Singapore (2004).

\bibitem{Boal} D. H. Boal, {\em Mechanics of the cell}, Cambridge
  University Press (2012).

\bibitem{Ambjorn_97} J. Ambj{\o}rn, B. Durhuus and T. Jonsson, {\em
  Quantum Geometry: A Statistical Field Theory Approach}, Cambridge
  University Press (1997).

\bibitem{Knizhnik_88} V. Knizhnik, A. M. Polyakov and
  A. B. Zamolodchikov, {\em Mod. Phys. Lett. A} {\bf 03}, 819 (1988).

\bibitem{Hammersley_65} J. M. Hammersley and D. J. A. Welsh,
  ``First-passage percolation, subadditive processes, stochastic
  networks and generalized renewal theory'', in {\em Bernoulli, Bayes,
    Laplace anniversary volume}, J. Neyman and L. M. LeCam eds.,
  Springer-Verlag (1965), p. 61.

\bibitem{Howard_04} C. D. Howard, ``Models of first passage
  percolation'', in {\em Probability on discrete structures},
  H. Keston ed., Springer (2004), p. 125.

\bibitem{Kesten_03} H. Kesten, {\em Prog. in Probability} {\bf 54}, 93
  (2003).

\bibitem{Eckhoff_13} M. Eckhoff, J. Goodman, R. van der Hofstad and
  F. R. Nardi, {\em J. Stat. Phys.} {\bf 151}, 1056 (2013).

\bibitem{Newman} M. E. J. Newman, {\em Networks: An Introduction},
  Oxford University Press (2010).

\bibitem{Lagatta_10} T. LaGatta and J. Wehr, {\em J. Math. Phys.} {\bf
  51}, 053502 (2010).

\bibitem{Lagatta_14} T. LaGatta and J. Wehr, {\em Comm. Math. Phys.}
  {\bf 327}, 181 (2014).

\bibitem{Krug_92} J. Krug and H. Spohn, ``Kinetic Roughening of
  Growing Surfaces'', in {\em Solids Far from Equilibrium},
  C. Godr\`eche ed., Cambridge University Press (1992), p. 479.

\bibitem{Barabasi} A.-L. Barab\'asi and H. E. Stanley, {\em Fractal
  Concepts in Surface Growth}, Cambridge University Press (1995).

\bibitem{Krug_97} J. Krug, {\em Adv. Phys.} {\bf 46}, 139 (1997).

\bibitem{Kardar_PRL86} M. Kardar, G. Parisi and Y.-C. Zhang, {\em
  Phys. Rev. Lett.} {\bf 56}, 889 (1986).

\bibitem{Chatterjee_13} S. Chatterjee, {\em Ann. Math.} {\bf 177}, 663
  (2013).

\bibitem{Auffinger_14} A. Auffinger and M. Damron, {\em Ann. Probab.}
  {\bf 42}, 1197 (2014).

\bibitem{Takeuchi_11} K. A. Takeuchi, M. Sano, T. Sasamoto and
  H. Spohn, {\em Sci. Rep.} {\bf 1}, 34 (2011).

\bibitem{Praehofer_02} M. Pr\"ahofer and H. Spohn, {\em
  J. Stat. Phys.} {\bf 108}, 1071 (2002).

\bibitem{Corwin_13} I. Corwin, J. Quastel and D. Ramenik, {\em
  Comm. Math. Phys.} {\bf 317}, 347 (2013).

\bibitem{Corwin_12} I. Corwin, {\em Random Matrices: Theor. Appl.}
  {\bf 1}, 1130001 (2012).

\bibitem{Alves_11} S. D. Alves, T. J. Oliveira and S. C. Ferreira,
  {\em Europhys. Lett.} {\bf 96}, 48003 (2011).

\bibitem{Oliveira_12} T. J. Oliveira, S. C. Ferreira and S. G. Alves,
  {\em Phys. Rev.} E {\bf 85}, 010601(R) (2012).

\bibitem{Takeuchi_PRL10} K. A. Takeuchi and M. Sano, {\em
  Phys. Rev. Lett.} {\bf 104}, 230601 (2010).

\bibitem{Yunker_13} P. J. Yunker, M. A. Lohr, T. Still, A. Borodin,
  D. J. Durian and A. G. Yodh, {\em Phys. Rev. Lett.} {\bf 110},
  035501 (2013); {\em ibid} {\bf 111} 209602 (2013).

\bibitem{Nicoli_13} M. Nicoli, R. Cuerno and M. Castro, {\em
  Phys. Rev. Lett.} {\bf 111}, 209601 (2013).

\bibitem{Sasamoto_PRL10} T. Sasamoto and H. Spohn, {\em
  Phys. Rev. Lett.} {\bf 104}, 230602 (2010).

\bibitem{Amir_CPAM11} G. Amir, I. Corwin and J. Quastel, {\em
  Comm. Pure Appl. Math.} {\bf 64}, 466 (2011).

\bibitem{Calabrese_11} P. Calabrese and P. Le Doussal, {\em
  Phys. Rev. Lett.} {\bf 106}, 250603 (2011).

\bibitem{Praehofer_PRL00} M. Pr\"ahofer and H. Spohn, {\em
  Phys. Rev. Lett.} {\bf 84}, 4882 (2000).

\bibitem{Praehofer_PA00} M. Pr\"ahofer and H. Spohn, {\em Physica} A
  {\bf 279}, 342 (2000).

\bibitem{Johansson_CMP00} K. Johansson, {\em Comm. Math. Phys.} {\bf
  209}, 437 (2000).

\bibitem{Dotsenko_JSM10} V. S. Dotsenko, {\em J. Stat. Mech.: Theory
  Exp.} {\bf 2010}, P07010 (2010).

\bibitem{Takeuchi_JSTAT12} K. A. Takeuchi, {\em J. Stat. Mech.: Theory
  Exp.} {\bf 2012}, P05007 (2012).

\bibitem{Johansson_00} K. Johansson, {\em Comm. Math. Phys.} {\bf
  209}, 437 (2000).

\bibitem{LSC_JSTAT11} J. Rodriguez-Laguna, S. N. Santalla and
  R. Cuerno, {\em J. Stat. Mech.: Theory Exp.} {\bf 2011}, P05032
  (2011).

\bibitem{SLC_PRE14} S. N. Santalla, J. Rodriguez-Laguna and R. Cuerno,
  {\em Phys. Rev.} E {\bf 89}, 010401(R) (2014).

\bibitem{Takeuchi.JSP.12} K. A. Takeuchi and M. Sano, {\em
  J. Stat. Phys.} {\bf 147}, 853 (2012).

\bibitem{Johansson.CMP.03} K. Johansson, {\em Comm. Math. Phys.} {\bf
  242}, 277 (2003).

\bibitem{Burns} K. Burns and M. Gidea, {\em Differential Geometry and
  Topology: with a view to dynamical systems}, CRC Press, Boca Rat\'on
  (2005).

\bibitem{Ferrari_JSP11} P. L. Ferrari and R. Frings,
  J. Stat. Phys. {\bf 144}, 1123 (2011).

\bibitem{Alves_JSTAT13} S. G. Alves, T. J. Oliveira and
  S. C. Ferreira, J. Stat. Mech. Theory Exp. {\bf 2013}, P05007
  (2013).

\bibitem{Alves_EPL11} S. G. Alves, T. J. Oliveira and S. C. Ferreira,
  {\em Europhys. Lett.} {\bf 96}, 48003 (2011).

\bibitem{Bornemann_MC10} F. Bornemann, {\em Math. Comput.} {\bf 79},
  871 (2010).

\bibitem{Tang92_etal} L.-H. Tang and H. Leschhorn, Phys. Rev. A {\bf
  45}, R8309 (1991); H. Leschhorn, Phys. Rev. E {\bf 54}, 1313 (1996).

\bibitem{Castellana} M. Castellana, A. Decelle and E. Zarinelli, {\em
  Phys. Rev. Lett.} {\bf 107}, 275701 (2011); M. Castellana, {\em
  Phys. Rev. Lett.} {\bf 112}, 215701 (2014).

\bibitem{Somoza_07} A. M. Somoza, M. Ortu\~no and J. Prior, {\em
  Phys. Rev. Lett.} {\bf 99}, 116602 (2007).

\bibitem{Git} {\tt http://github.com/jvrlag/riemann}.

\end{thebibliography}
\end{document}